\documentclass{aa}
 
\usepackage{H0594_AAcite,graphics,times}
 
\begin{document}
 
\thesaurus{05         
          (08.12.1;           
           08.12.2;           
           08.12.3;           
           10.15.2 Pleiades)}  

\title{Brown Dwarfs in the Pleiades \thanks{Based on observations collected at the Nordic Optical Telescope (NOT), La Palma}}

\subtitle{II. A deep optical and near infrared survey}

\author{L. Festin}

\offprints{L. Festin}

\institute{Astronomical Observatory,
	   Box 515,
	   S-751 20 UPPSALA,
           Internet: leif@astro.uu.se
           }

\date{Received            ; accepted           }

\maketitle

\begin{abstract} 
The brown dwarf population in the Pleiades cluster has been probed in a deep 850 arcmin$^{2}$ $RIJK$ survey. The survey is complete to $I=21.4$ in 76\,\% of the area and to $I=20.2$ in the remaining 24\,\%. Photometry of 32 previously known members is presented together with 8 new candidates, four of which are below the brown dwarf limit. The faintest one is the lowest mass brown dwarf candidate found hitherto in the Pleiades ($I=20.55$, $0.04$ M$_{\odot}$). The derived Pleiades luminosity function is compared to the most recent theoretical mass-luminosity relations and is consistent with a power-law index in the mass function between 0 and 1 to the limit of this survey.
   \end{abstract}

\keywords{Stars: late-type - low-mass, brown dwarfs - luminosity function, mass function - open clusters and associations:individual: Pleiades}
%

\section{Introduction}
The Pleiades cluster has been the major target of several recent surveys for brown dwarfs (BDs)  (Hambly et al. 1993 (HHJ); Jameson \& Skillen 1989 (JS); Schultz 1997; Stauffer et al. 1989, 1994a; Williams et al. 1996; Zapatero Osorio 1997; Zapatero Osorio et al. 1997a, b (ZRM, ZMR); Festin 1997; Cossburn et al. 1997). Its nearness ($116 \pm 3$ pc, Mermilliod et al. \cite*{mermilliod97}) and youth (120 Myr, Basri et al. \cite*{basri96}) makes the rapidly cooling BDs still rather bright and easy to detect. The first bona fide Pleiades BD was reported by Rebolo et al. \cite*{rebolo95}. This object, known as Teide1, passed the lithium test \cite{rebolo96} and should thereby have a maximum mass of 0.06 M$_{\odot}$. By now, on the order of 10 BDs have been confirmed in the Pleiades.

The present observations were designed to probe low-mass BDs in the Pleiades and to provide accurate first-epoch data for future proper motion determination. This paper is an extension of the $IJK$ survey described in Festin \cite*{festin97a}.

\begin{figure}
\resizebox{\hsize}{!}{\includegraphics{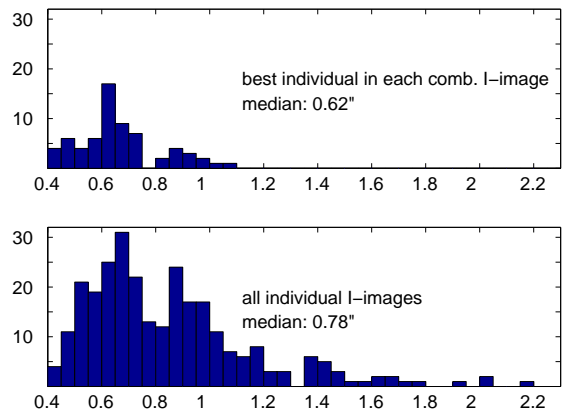}}
\caption{Histogram of seeing in the $I$ images}
\label{Figseeing}
\end{figure}

\section{Observations and Reductions}
For a Teide1-type object, the flux rises from 10 to almost 100\,\% of its peak value between $I$ and $J$. Therefore, although providing a rather short wavelength baseline, $I-J$ was chosen as the primary temperature indicator. Its efficiency is proved by the large gap between the Pleiades sequence and the background stars in the $I$ vs $I-J$ diagram. 
Most observational effort was spent on obtaining high-quality images in $I$ (Fig.~\ref{Figseeing}) to get good first-epoch coordinates and to clean out galaxies. To minimize the effects of nonuniform pixels and rotation angle, each $I$ field was observed at four different field angles (0, 90, 180 and 270 degrees). Complementary photometry was taken in $RJK$.\\ 
 
\begin{figure}[h]
\resizebox{\hsize}{!}{\includegraphics{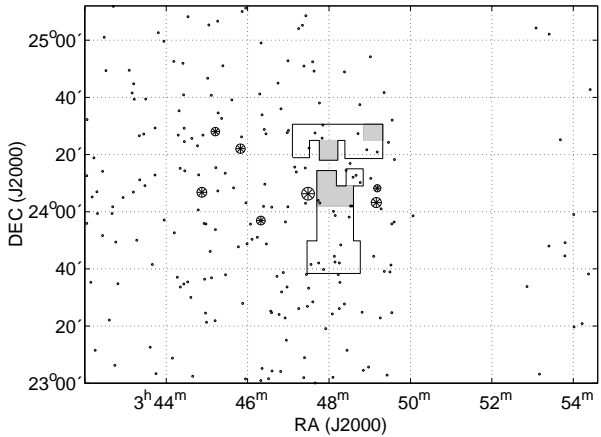}}
\caption{The observed area ($RI+IJ$) superposed on proper motion members from HHJ and the "Seven Sisters". The shaded area outlines the region covered only in $RI$. The empty vertical strip is the region where the first and second epoch plates used by HHJ don't overlap}
\label{Figfield}
\end{figure}

All observations were carried out at the 2.5\,m Nordic Optical Telescope (NOT), La Palma. The survey covers 648 arcmin$^{2}$ in $IJ$ (of which 240 arcmin$^{2}$ were also covered in $K$) + 200 arcmin$^{2}$ in $RI$ only. The total area is 848 arcmin$^{2}$ near the centre of the Pleiades (Fig.~\ref{Figfield}). A summary of the observations is given in Table~\ref{Tabobsoverview}. 

\begin{table}[h]
\caption[]{Observations}
\label{Tabobsoverview}
\begin{flushleft}
\begin{tabular}{lllll}
\hline
date & filter  & arcmin$^{2}$ & instrument & scale  \\
     &         &              &            & "/px   \\
\hline
1995 Aug30-Sep09 & $JK$     & 240  & ARNICA  & 0.56\\
1995 Nov11-Nov21 & $I$      & 468  & BROCAM1 & 0.18\\
1996 Sep23-Sep27 & $J$      & 408  & ARNICA  & 0.56\\
1996 Nov04-Nov07 & $R$      & 200  & ALFOSC  & 0.19\\
1996 Nov04-Nov07 & $I$      & 396  & ALFOSC  & 0.19\\
\hline
\end{tabular}
\end{flushleft}
\end{table}

The ARNICA is a 256x256 NICMOS3 near-IR array. Due to the high density of bad pixels each final image was derived as the median of 5--7 slightly dithered subimages. BROCAM1 is a CCD camera equipped with a Tek1K chip, and ALFOSC is a combined focal reducer/low--resolution spectrograph equipped with a Loral2K chip.\\

\subsection{Reductions and photometric calibrations}
The reductions were carried out within IRAF (Image Reduction and Analysis Facility) \footnote{IRAF is distributed by National Optical Astronomy Observatories (NOAO), which is operated by the Association of Universities for Research in Astronomy, Inc., under contract with the National Science Foundation.}. All science frames were bias and flatfield corrected in a standard fashion.

The transformation to the Kron-Cousins system in $RI$ was based on standard stars selected from Landolt \cite*{landolt92}. Since the target objects are very red, care was taken to include the reddest dwarf stars from this list.

On a typical photometric night 10 different standard fields were observed, each containing 3 stars on the average. The colour range covered by the standard stars was $0 < R-I < 2.2$, the red limit defined by G45-20 (M6V), the reddest dwarf star in Landolt's list. G3-33 (M5V) and G44-40 (M4V) were also used. Transformation equations with linear colour terms were derived in a standard fashion, resulting in standard star residuals of less than 0.02 mag across the whole colour range. The $I-J$ colours of our reddest targets indicate spectral types of $\sim$\,M9V, beyond the standard star range by $\sim$ 0.3 mag in $R-I$. This is not a serious problem here, since even for a 50\,\% change in the colour coefficient, the resulting $I$ magnitude would not be offset by more than 0.1 mag from the correct value in the standard system.

In $JK$, standard stars provided by the ARNICA team were used \cite{hunt95,casali92}. The Hunt transformation from ARNICA to the CIT system was adopted: $J_\mathrm{CIT} = J_\mathrm{ARNICA}$, $K_\mathrm{CIT} = K_\mathrm{ARNICA}+0.12$. The errors in these transformations are $\sim$ 0.05 mag rms. Night-to-night shifts in the zero point of up to 0.20 mag were noticed during the two ARNICA runs. The errors in these shifts are $\sim0.05$ mag rms. Adding these two errors we end up with a final $1\,\sigma$ calibration error in the $JK$ photometry of 0.07 mag.

The colour correction for $I$ in the $IJ$ fields was done by transforming $I-J$ (using a zero-point corrected $I$) to $R-I$ via relations in Leggett \cite*{leggett92}. This procedure induced an extra error of 0.03 mag, and the final $1\,\sigma$ transformation error in $I$ is 0.04 mag.

The extraction of instrumental magnitudes in the science fields was done with an empirical growth-curve technique outlined in Festin \cite*{festin97a}.

\subsection{Completeness limits}
\label{completenesslimit}

\begin{figure}
\resizebox{\hsize}{!}{\includegraphics{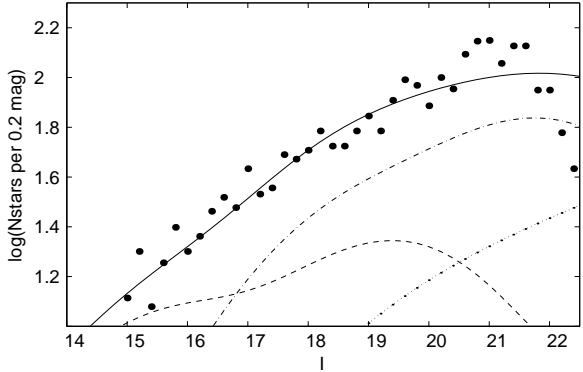}}
\caption{The observed $I$ counts (dots) compared to the model (solid line) consisting of a halo (dotted line), thin (dashed line) and thick disk (dot-dashed line) added together. For details, see Sect.~\ref{completenesslimit}}
\label{Figcompleteness}
\end{figure}

The completeness limit was defined as the magnitude at which log\,($N_\mathrm{stars}$) vs $I$ deviates from a straight line. This is justified by the model predictions in Fig.~\ref{Figcompleteness}, and by star counts in Santiago et al. \cite*{santiago96}, increasing to $I=23.5$, well beyond our limit. 

The luminosity functions used in the model were taken from Gould et al. \cite*{gould97} ($M_\mathrm{V}>8$) and Scalo \cite*{scalo86} ($M_\mathrm{V}<8$). The halo contribution was estimated from the model in Bahcall \& Soneira \cite*{bahcall80}, using an axis ratio of $c/a = 0.6$ and a local normalization of 1/500 of the local disk density. The disk model consisted for stars fainter than $M_\mathrm{V}=5$ of two components, with scale heights and normalizations taken from Gould et al. \cite*{gould97}, 700 pc (Gould upper limit), 22\,\% and 320 pc, 78\,\% respectively. The bright stars ($M_\mathrm{V}<5$) were modelled by a disk of scale height 250 pc. A scale length of 3.5 kpc was adopted for all disk cases.

The completeness limit should be set as the magnitude at which the $I$ counts start to decrease, which for our data occurs approximately at the same point as the deviation from a straight line.   
The completeness limits for the whole survey as defined by the worst cases are $I = 21.4$,  $J = 18.8$ in the $IJ$ part, and $I = 21.4$, $R = 22.3$  ($R = 22.3 \cor I\sim20.2$ on the Pleiades sequence) in the $RI$ part. For individual fields the internal magnitude error at the completeness limit is $\sim$ 0.1 mag.
  
\subsection{Classification of objects}
The best subframe of each combined $I$ image was used to identify stars and binaries. Approximately 15000 sources were classified by eye as stars, binaries, galaxies or too faint for classification. The final $IJ$ sample consists of 1513 stars brighter than the completeness limits in both filters. The corresponding number for $RI$ is 693.

Binaries and stars close to galaxies were checked by point-spread function fitting in addition to the routine procedure \cite{festin97a}. Binaries that were resolved in $I$ but remained unresolved in the other filter ($R$, $J$ or $K$) were considered as unresolved systems in the colour-magnitude diagrams.

\section{Extraction of the Pleiades candidates}
The distance modulus of the Pleiades was adopted from Mermilliod et al. (1997), $m-M=5.32$. Uncertainty in the mean distance and internal spread cause a $1\,\sigma$ error in the absolute magnitude for a single member of $\sim0.1$ mag. The mean colour excess in this area, $E_\mathrm{B-V}=0.04$ \cite{stauffer89}, gives via relations in Winkler \cite*{winkler97} $A_\mathrm{I}=0.07$ and $E_\mathrm{I-J}=0.04$.

\begin{figure}
\resizebox{\hsize}{!}{\includegraphics{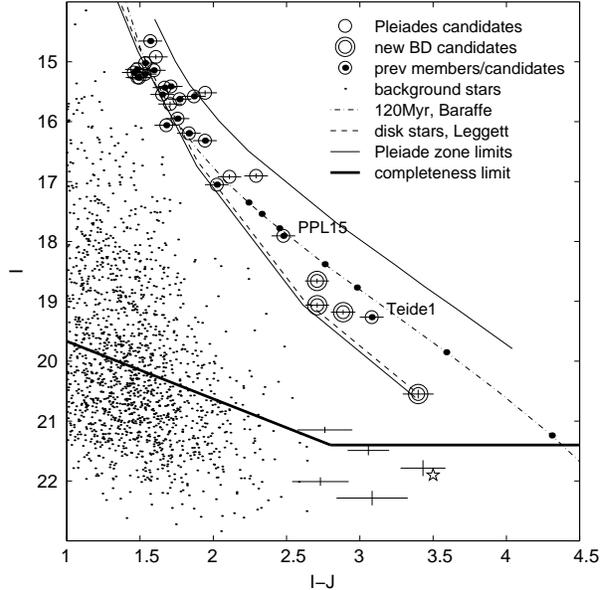}}
\caption{Colour-magnitude diagram for all stars detected in both $I$ and $J$ in 648 arcmin$^{2}$. Ticks (M$_{\odot}$): 0.035, 0.045, 0.055, 0.060, 0.070, 0.075, 0.080 \protect\cite{baraffe98}. The naked error bars show the field VLMSs. The pentagram is GD165B at the Pleiades distance and reddening. The error bars are $1\,\sigma$ including internal and transformation errors} 
\label{FigIJ}
\end{figure}

The area in the $I$ vs $I-J$ diagram (Fig.~\ref{FigIJ}) which the Pleiades should occupy is defined by the most recent evolutionary model of I. Baraffe \cite{baraffe98} and a sequence of late disk stars \cite{leggett92} including  extinction, colour excess and distance error as defined above. The bright part of the Pleiades is well fitted, but as no present model is able to fit the colours of dwarfs later than M6 \cite{allard97}, we choose to define the faint limit for $I-J>2$ by the empirical disk sequence . 

The question of age spread is unclear and has not been taken into account. For a discussion see Stauffer et al. \cite*{stauffer95}. The bright limit of the Pleiades zone is the binary envelope, offset from the single-star sequence by 0.75 mag.

The 29 objects within the defined limits (Fig.~\ref{FigIJ}) are listed in Table~\ref{Tabphotometry} together with additional Pleiades members that were saturated in $I$ or not measured in $J$. The five very red objects at the bottom are possible very-low-mass stars (VLMSs) or even BDs in the field, and also of great interest. NPL43 deserves special attention, since it fits in as a GD165B-type object in the Pleiades ($I=21.9$, $I-J=3.5$, Jones et al. \cite*{jones94}). 
 
\begin{figure*}
\resizebox{\hsize}{!}{\includegraphics{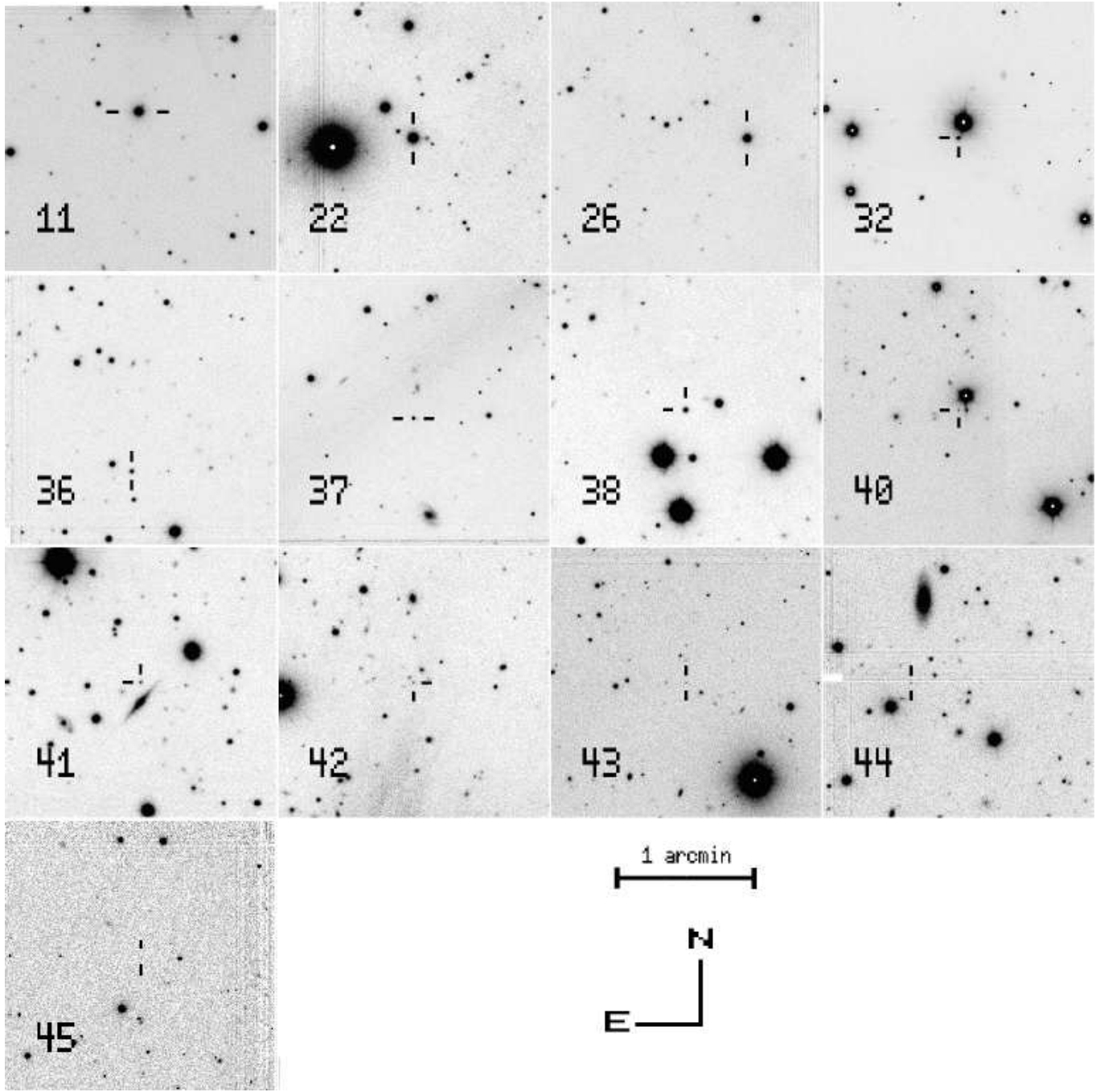}}
\caption{Finding charts for the new likely Pleiades members and the field VLMSs in Table~\ref{Tabphotometry}}
\label{Figfindcharts}
\end{figure*}

Finding charts are provided in Fig.~\ref{Figfindcharts} for all Pleiades candidates not previously published and for the five faintest objects in Table~\ref{Tabphotometry}. 

No additional candidates were found in the 200 arcmin$^{2}$ $RI$ fields. This is  consistent with the $IJ$ findings and is due to the smaller area covered and the shallower depth in the photometry.

All previously known Pleiades in the area were detected and recognized as photometric members, a good reliability test of this survey.
  
\subsection{Wide field binaries?}
Boxes of 10''x10'' were centred on all objects in Table~\ref{Tabphotometry} and searched for companions. Additional stellar sources were found in 6 cases (see Table~\ref{Tabphotometry}). For NPL29 and 37 (marked with ? in Table~\ref{Tabphotometry}) some signal is seen, but it is not clear whether it arises from a galaxy or a star. By comparing these counts to the field background we conclude that there is no statistically significant excess of stars within the boxes.\\

\begin{table*}
\caption[]{Photometry of Pleiades members and new candidates. NPL in the first column indicates that the photometric Pleiades candidates were identified in a survey carried out at the NOT. In the last column, A means a single star , B and C one or two possible wide-field companions. A* marks the possible unresolved binaries. The seeing in the best image of each object is also given. A pm$\pm$ indicates if the proper motion derived by Hambly (priv. comm.) is consistent ($+$) or not ($-$) with membership. The photometric errors are internal and should be added to the transformation errors (0.04 mag in $I$ and 0.07 mag in $JK$) to get the true formal error. Coordinates are accurate to $\sim 1\arcsec$. HHJ: Hambly et al. \cite*{hambly93}, WILL: Williams et al. \cite*{williams96}, JS: Jameson \& Skillen \cite*{jameson89}, PPL: Stauffer et al. \cite*{stauffer89,stauffer94a}, NOT: Festin \cite*{festin97a}
}
\label{Tabphotometry}
\begin{flushleft}
\begin{minipage}[t]{10cm}
\begin{tabular}{llllllll}
\hline
NPL & $I_{\rm KC}$  & $J_{\rm CIT}$ & $K_{\rm CIT}$ & RA (J2000) & DEC (J2000) & prev id & comment \\
\hline
 1 &   sat       & 12.78 (0.01)&   -         & 3:48:39.84 & 24:12:44.0 & HHJ347       &\makebox[0.8 cm][l]{A}   0.63" \\  
 2 &   sat       & 12.51 (0.01)& 11.68 (0.01)& 3:47:50.78 & 24:30:20.3 & HHJ389       &\makebox[0.8 cm][l]{A}   0.61" \\  
 3 &   sat       & 12.35 (0.01)&   -         & 3:47:30.60 & 24:22:15.3 & HHJ408       &\makebox[0.8 cm][l]{ABC} 0.72" \\  
 4 &   sat       & 13.56 (0.01)& 12.46 (0.01)& 3:47:39.33 & 24:27:33.3 & HHJ272       &\makebox[0.8 cm][l]{A}   0.63" \\  
 5 &   sat       & 13.25 (0.01)& 12.36 (0.01)& 3:48:07.91 & 23:44:25.2 & HHJ288       &\makebox[0.8 cm][l]{A}   0.46" \\  
 6 &   sat       & 12.74 (0.01)& 12.00 (0.01)& 3:48:15.38 & 23:42:06.9 & HHJ336       &\makebox[0.8 cm][l]{A}   0.62" \\  
 7 &   sat       &   sat       & 10.99 (0.01)& 3:47:33.54 & 23:41:29.8 & HHJ424       &\makebox[0.8 cm][l]{A}   0.56" \\  
 8 &   sat       & 12.96 (0.01)&   -         & 3:49:10.97 & 24:20:52.1 & HHJ287       &\makebox[0.8 cm][l]{A}   0.87" \\  
 9 & 13.45 (0.01)&   -         &   -         & 3:47:46.44 & 24:03:02.8 & HHJ438       &\makebox[0.8 cm][l]{A}   1.12" \\  
10 & 14.65 (0.01)& 13.08 (0.01)&   -         & 3:48:25.14 & 24:14:25.0 & HHJ314       &\makebox[0.8 cm][l]{AB}  0.92" \\  
11 & 14.92 (0.01)& 13.31 (0.01)&   -         & 3:48:13.30 & 23:58:46.8 &              &\makebox[0.8 cm][l]{AB}  0.84", pm+ \\  
12 & 15.02 (0.01)& 13.49 (0.01)&   -         & 3:48:06.63 & 24:00:07.5 & HHJ240       &\makebox[0.8 cm][l]{A}   0.92" \\  
13 & 15.10 (0.01)& 13.65 (0.01)&   -         & 3:48:09.22 & 23:58:40.1 & HHJ225       &\makebox[0.8 cm][l]{A}   0.84" \\  
14 & 15.14 (0.01)& 13.55 (0.01)& 12.82 (0.01)& 3:47:44.70 & 23:42:01.8 & HHJ152       &\makebox[0.8 cm][l]{A}   0.51" \\  
15 & 15.18 (0.01)& 13.72 (0.01)&   -         & 3:48:46.06 & 24:10:14.6 & HHJ207       &\makebox[0.8 cm][l]{AB}  0.58" \\  
16 & 15.20 (0.01)& 13.67 (0.01)& 12.98 (0.01)& 3:47:49.84 & 24:25:44.0 & HHJ202       &\makebox[0.8 cm][l]{A}   0.58" \\  
17 & 15.22 (0.01)&   -         &   -         & 3:49:27.53 & 24:24:14.4 & HHJ192       &\makebox[0.8 cm][l]{AB}  0.91" \\  
18 & 15.26 (0.01)& 13.78 (0.01)& 12.95 (0.01)& 3:48:17.15 & 23:48:25.5 & HHJ188       &\makebox[0.8 cm][l]{A}   0.48" \\  
19 & 15.26 (0.01)& 13.77 (0.01)& 12.90 (0.01)& 3:48:08.99 & 23:42:25.3 & HHJ156       &\makebox[0.8 cm][l]{A}   0.65" \\  
20 & 15.41 (0.01)& 13.70 (0.01)&   -         & 3:48:31.69 & 24:02:01.2 & HHJ197       &\makebox[0.8 cm][l]{A}   0.71" \\  
21 & 15.43 (0.01)& 13.76 (0.01)&   -         & 3:48:33.63 & 24:02:01.6 & HHJ184       &\makebox[0.8 cm][l]{A}   0.71" \\  
22 & 15.52 (0.01)& 13.57 (0.01)&   -         & 3:47:15.38 & 24:23:30.8 &              &\makebox[0.8 cm][l]{A*}   0.67", pm+ \\  
23 & 15.55 (0.01)& 13.90 (0.01)& 13.13 (0.01)& 3:47:52.11 & 23:39:48.2 & HHJ122       &\makebox[0.8 cm][l]{A}   0.50" \\  
24 & 15.58 (0.01)& 13.71 (0.01)&   -         & 3:48:32.64 & 23:52:41.3 & WILL1        &\makebox[0.8 cm][l]{A}   0.91", pm- \\  
25 & 15.62 (0.01)& 13.86 (0.01)&   -         & 3:48:29.78 & 23:58:07.8 & HHJ132,WILL3 &\makebox[0.8 cm][l]{A}   0.72" \\  
26 & 15.70 (0.01)& 14.00 (0.01)&   -         & 3:47:07.81 & 24:23:36.6 &              &\makebox[0.8 cm][l]{A}   0.67", pm+ \\  
27 & 15.95 (0.01)& 14.19 (0.01)&   -         & 3:48:35.46 & 24:12:03.6 & HHJ96        &\makebox[0.8 cm][l]{AB}  0.67" \\  
28 & 16.06 (0.01)& 14.38 (0.01)&   -         & 3:47:07.77 & 24:21:39.0 & JS9          &\makebox[0.8 cm][l]{A}   0.79", pm- \\  
29 & 16.19 (0.01)& 14.36 (0.01)&   -         & 3:48:42.65 & 24:27:20.5 & HHJ44,WILL6  &\makebox[0.8 cm][l]{AB?} 0.91" \\ 
30 & 16.32 (0.01)& 14.37 (0.01)&   -         & 3:48:10.15 & 23:59:19.8 & PPL12        &\makebox[0.8 cm][l]{A}   0.83", pm?\\  
31\footnote{$R_{KC}=18.21$ (0.04)}    & 16.69 (0.01)&   -         &   -         & 3:47:44.09 & 24:03:56.8 & HHJ26        &\makebox[0.8 cm][l]{A}   1.12", phot. nonmember\\  
32 & 16.90 (0.01)& 14.61 (0.01)& 13.58 (0.01)& 3:48:03.61 & 23:44:13.1 & NOT1         &\makebox[0.8 cm][l]{A*}   0.61", pm? \\  
33 & 16.92 (0.01)& 14.81 (0.01)&   -         & 3:48:23.60 & 24:22:35.7 &              &\makebox[0.8 cm][l]{A}   0.95", pm- \\  
34 & 17.05 (0.01)& 15.03 (0.01)&   -         & 3:48:55.68 & 24:21:41.0 & HHJ8         &\makebox[0.8 cm][l]{A}   0.65" \\  
35 & 17.91 (0.01)& 15.43 (0.01)& 14.48 (0.02)& 3:48:04.82 & 23:39:32.0 & PPL15        &\makebox[0.8 cm][l]{A*}   0.63" \\  
36 & 18.66 (0.01)& 15.95 (0.02)& 15.12 (0.02)& 3:48:19.07 & 24:25:15.0 &              &\makebox[0.8 cm][l]{A}   0.65" \\  
37 & 19.06 (0.01)& 16.36 (0.02)&   -         & 3:47:12.06 & 24:28:31.4 &              &\makebox[0.8 cm][l]{AB?} 0.63" \\  
38 & 19.18 (0.01)& 16.30 (0.03)&   -         & 3:47:50.37 & 23:54:48.6 &              &\makebox[0.8 cm][l]{A}   0.87" \\  
39 & 19.26 (0.01)& 16.18 (0.01)&   -         & 3:47:17.90 & 24:22:31.9 & Teide1       &\makebox[0.8 cm][l]{A}   0.67" \\  
40 & 20.55 (0.07)& 17.15 (0.03)&   -         & 3:48:49.12 & 24:20:25.4 &              &\makebox[0.8 cm][l]{A}   0.62" \\  
41 & 21.15 (0.04)& 18.39 (0.17)&   -         & 3:48:04.73 & 23:51:02.3 &              &\makebox[0.8 cm][l]{A}   0.93" \\  
42 & 21.49 (0.07)& 18.43 (0.10)&   -         & 3:48:32.32 & 24:13:18.5 &              &\makebox[0.8 cm][l]{A}   0.70" \\  
43 & 21.79 (0.13)& 18.36 (0.04)& 16.71 (0.09)& 3:48:27.36 & 23:46:20.3 & NOT3         &\makebox[0.8 cm][l]{A}   0.46" \\  
44 & 22.01 (0.06)& 19.28 (0.17)&   -         & 3:48:19.48 & 23:56:25.7 &              &\makebox[0.8 cm][l]{A}   0.66" \\ 
45 & 22.29 (0.12)& 19.20 (0.20)& 17.38 (0.21)& 3:47:33.15 & 23:49:12.8 & NOT2         &\makebox[0.8 cm][l]{A}   0.48" \\  

\end{tabular}
\end{minipage}
\end{flushleft}
\end{table*}

\section{Contamination}
Unresolved galaxies, background giants and field M dwarfs are the main possible contamination sources of the Pleiades candidates. The galaxy/star separation is believed to be reliable to the completeness limit and is not discussed further. 

\subsection{Giants}
Consider the region $I>17$ and $I-J>2$ ($V-I\ga3.5$) in Fig.~\ref{FigIJ}. The light from such apparently faint giants passes through approximately all the interstellar matter in this direction ($A_{V}\sim0.6$ \cite{burstein82}, $E_{V-I}\sim0.25$ \cite{winkler97}). Thus, only giants redder than $V-I=3.2$ are left as possible contaminators, i.e. red giants brighter than $M_{V}=-2$. We adopt here the conservative limit $M_{V}=0$, and consider all brighter stars in the halo and disk luminosity functions (LFs) in Fig. 12 of Bahcall \cite*{bahcall86} as red giants. From the disk and halo models in the same paper by Bahcall, the number of contaminators is estimated to less than 0.1 in our total field. Thus it is highly improbable that the Pleiades BD region is contaminated by giants.

\subsection{M dwarfs}
The Galaxy model used for the completeness limit (Sect.~\ref{completenesslimit}) was also used to check the possible M-dwarf contamination. Interstellar reddening was included as a dust component of scale height 100 pc \cite{bahcall80}, normalized by the total galactic extinction in the Pleiades direction ($A_{I}\sim0.35$, $E_{I-J}\sim0.2$ \cite{burstein82,winkler97}).

\begin{figure}
\resizebox{\hsize}{!}{\includegraphics{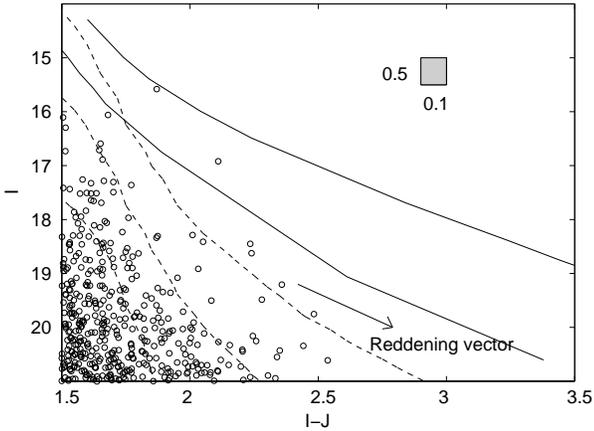}}
\caption{M-dwarf contamination. The thin dashed lines show the M-dwarf isonumbers. The contours correspond to 0.1, 1 and 10 stars per indicated square respectively in the 648 arcmin$^{2}$ $IJ$ part of the survey. The solid lines are the Pleiades zone as defined in Fig.~\ref{FigIJ}. The rings are our background stars}
\label{Figcontamination}
\end{figure}

The photometric LF in Gould et al. \cite*{gould97} was transformed  to $I$ and $I-J$ via relations in Leggett \cite*{leggett92} and scaled appropriately. The number of stars  per magnitude ($I$) and colour interval ($I-J$) was integrated to the magnitude limit of this survey. The derived M-dwarf isonumbers are compared to our data in Fig.~\ref{Figcontamination}. 
The conclusion is that it is not likely that the proposed new Pleiades BDs are field M dwarfs. 

\section{Results}
Eight new Pleiades candidates have been identified, four of which are possible BDs. Three of the four brightest new candidates have proper motions consistent with Pleiades membership (Hambly, priv. comm.). Two probable members (NPL22 \& 32) stick out from the single-star sequence and are analyzed as binaries together with the spectroscopic binary PPL15 (NPL35). 
A number of faint very red objects were also found. Two of those were measured also in $K$ and show colours similar to GD165B and are possible field BDs.

\subsection{Overall appearance}
\begin{figure}
\resizebox{\hsize}{!}{\includegraphics{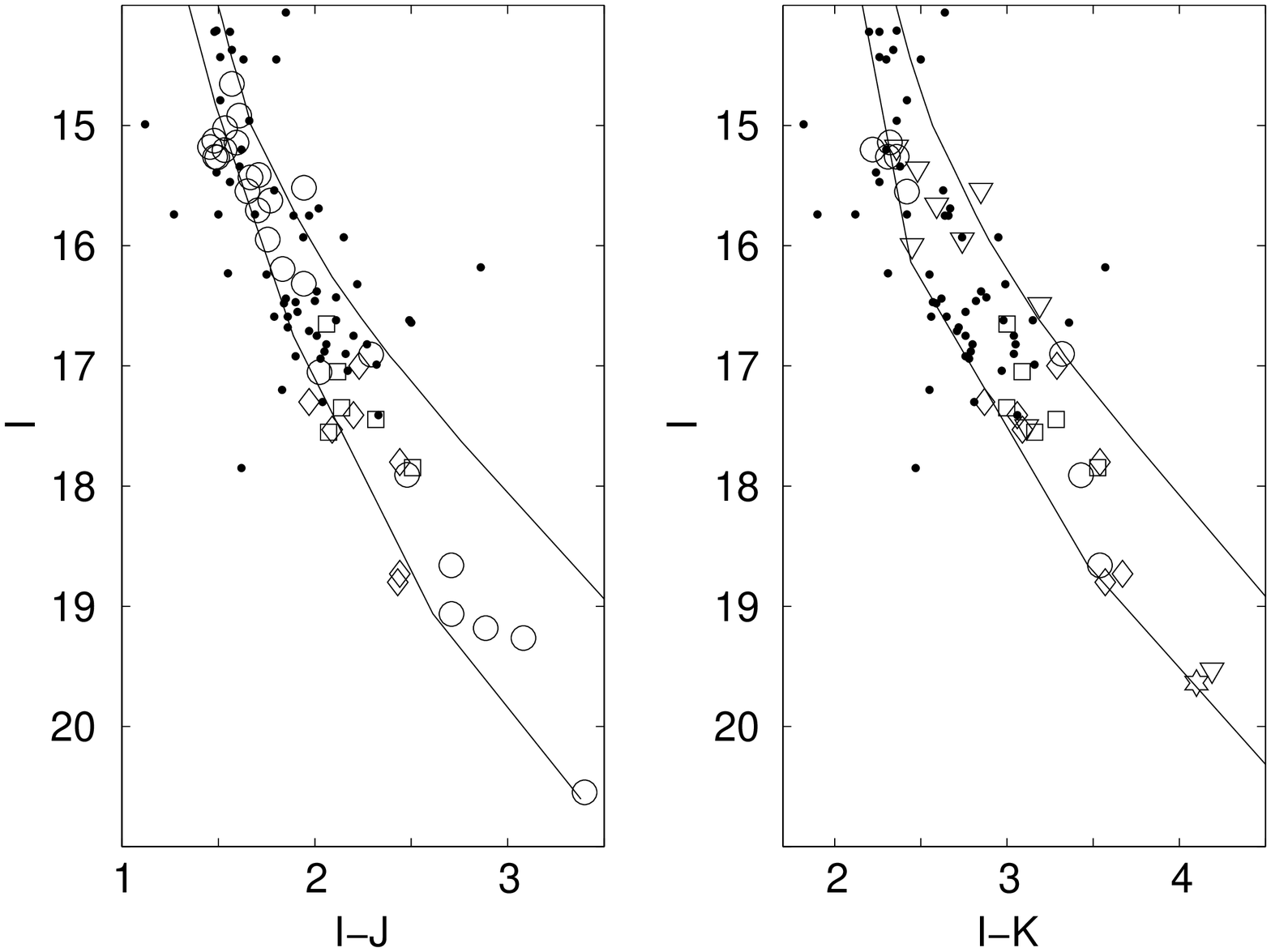}}
\caption{A comparison to other recent surveys. (squares: Stauffer et al.  \protect\cite*{stauffer89,stauffer94a}, dots: HHJ-stars in Steele et al.  \protect\cite*{steele93,steele95}, triangles: Williams et al. \protect\cite*{williams96}, diamonds: Zapatero Osorio et al. \protect\cite*{zapatero97b}, open rings: this survey, star: PIZ1, Cossburn et al.  \protect\cite*{cossburn97}. The solid lines are the Pleiades zone as defined in Fig.~\ref{FigIJ}}
\label{Figcomparison}
\end{figure}

In Fig.~\ref{Figcomparison} this survey is compared to several other recent surveys. Known nonmembers have been excluded. 
The dispersion of the Steele et al. \cite*{steele93,steele95} data can probably be explained by photometric uncertainty, since most of their $I$ magnitudes are  photographic.
Note that the faint Pleiades sequence is slightly bluer than the Baraffe et al. \cite*{baraffe98} model. Part of this may be due to incomplete line lists and not yet included dust formation in the models. Note also that Mermilliod et al. \cite*{mermilliod97} found from Hipparcos data that the Pleiades cluster is peculiar in the sense that its main sequence is $\sim0.4$ mag fainter than other nearby clusters, such as the Hyades and Praesepe. The reason for this peculiarity is not known, and may also hide part of the model deviation.

\subsection{Individual objects}
\label{individual}
The objects in Table~\ref{Tabphotometry} that are of special interest are individually discussed and compared to other papers below.

{\bf NPL11}, {\bf 22} and {\bf 26} have proper motions consistent with membership (Hambly, priv. comm.), although not present in HHJ. NPL22 is also a possible binary, best fitted by two components of equal brightness, $I_\mathrm{A}=I_\mathrm{B}=16.3$ 

{\bf NPL24}, {\bf 28} and {\bf 33} have proper motions that are not consistent with membership.

{\bf NPL30} (PPL12) and {\bf 32} have uncertain proper motions. NPL30 has a radial velocity consistent with membership \cite{stauffer94b}. NPL32 is very close to a bright star, which due to blending makes the photographic proper motion uncertain. If NPL32 is a member, its position above the Pleiades sequence indicates an unresolved binary, best fitted by $I_\mathrm{A}=17.4$, $I_\mathrm{B}=18.0$.

{\bf NPL31} (HHJ26) is, as also found by Steele et al. \cite*{steele93}, an $RI$ nonmember.

{\bf NPL35} (PPL15) has been measured by several authors recently (Stauffer et al. 1994a; Basri et al. 1996; ZMR),  and also found to be a spectroscopic binary \cite{basri97}. 
From the primary component's possible locii in our colour-magnitude diagrams, the secondary's mass is $\sim0.03$ M$_{\odot}$, consistent with ZMR. The $I$ magnitudes would be $I_\mathrm{A}=17.96$ \& $I_\mathrm{B}=21.3$. A heavier and brighter secondary would force the primary below the disk sequence. 

{\bf NPL36-38} are all below the BD limit. NPL37 shows a slight brightness enhancement at the edge of the stellar profile. It is not clear wether this is a background star or galaxy or if NPL37 itself is a compact galaxy.

{\bf NPL39} (Teide1). Our result is $I=19.26$, $J=16.18$. The values given in ZMR and ZRM are $I=18.80$ and $J=16.37$. The $J$ magnitude agrees fairly well, but the difference in $I$ ($0.46$ mag) is clearly exceeding the error-bar limits. Teide1 is present on the same $I$ frame as JS4 \& 9 in the RGO La Palma archive. Our magnitudes for JS4 \& 9 agree within $0.03$ mag with JS. The relative magnitude offset to Teide1 in the archive frame gave $I=19.3$, in good agreement with our $I=19.26$. Since two of the measurements give almost identical magnitudes and the third deviates, it is likely that the explanation to the discrepancy sits in the photometry of ZRM and not in intrinsic variability. This conclusion is supported by the $J$ magnitude staying roughly constant.

{\bf NPL40} would be the lowest mass BD found so far in the Pleiades ($\sim0.04$ M$_{\odot}$) if it can be confirmed as a member.

{\bf NPL41-45} are probably not Pleiades members, but still of interest. NPL43 \& 45 have $K$ magnitudes which place them as GD165B-type objects, i.e. possible field BDs.\\

All the JS candidates except JS9, fall outside our Pleiades limits, confirming the result of ZRM. JS9 itself has a proper motion inconsistent with membership \cite{hambly91}.

\subsection{The luminosity function and the mass function}
\label{lumfunc}
\begin{figure}
\resizebox{\hsize}{!}{\includegraphics{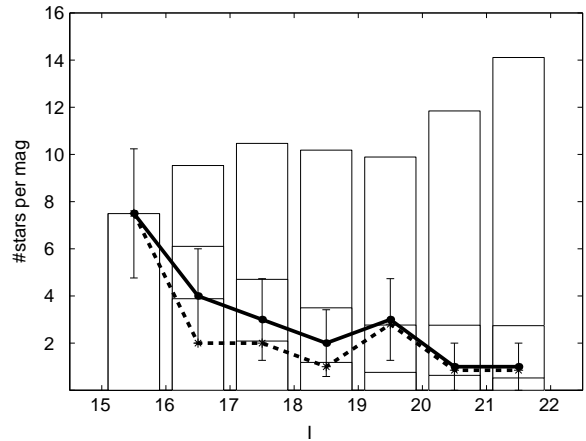}}
\caption{The LF from this survey (Table~\ref{Tabimf}). The solid and dashed lines show \emph{Case 1} and \emph{2} respectively (Sect.~\ref{lumfunc}). The error bars are only included for \emph{Case 1} and are Poissonian, defined by the number of stars in each bin. The bars are the LFs derived for MF indices 2 (uppermost), 1 (mid) and 0 (lowest). The incompleteness in the two last bins have not been corrected for }
\label{Figimf}
\end{figure}

In Fig.~\ref{Figimf} we show the derived Pleiades LF. The included objects are listed in Table~\ref{Tabimf}. 

\emph{Case 1} in Fig.~\ref{Figimf} treats the proposed binaries NPL22 and 32 as separate components, with magnitudes as in Sect.~\ref{individual}. \emph{Case 2} rejects NPL22 and 32 as nonmembers, since if they are not binaries they are too bright to be members. Both \emph{Case1} \& \emph{2} treat PPL15 as two separate components.

The theoretical LF was derived from evolutionary models in Baraffe et al. \cite*{baraffe98}, which are based on the latest generation of non-grey atmosphere models \cite{allard97}. 
Three different MF-indices were considered , $n=0$, 1 and 2 ($n$ is defined by $dN=const*m^{-n}dm$, $m$ = mass and $dN$ = the number of stars per mass interval $dm$). We normalized to 7.5 stars in the interval $15 < I_\mathrm{KC} < 16$, the mean density of HHJ stars (corrected for the 80\,\% completeness estimated in HHJ) in this part of the Pleiades. Our fields should not be used for the normalization, since they were selected to contain HHJ stars.\\

\begin{table}
\caption[]{The luminosity function}
\label{Tabimf}
\begin{flushleft}
\begin{tabular}{lll}
\hline
$I$ interval & included objects  & \# objects/bin  \\
\hline
$15-16$ & mean surface density from HHJ  & 7.5 \\
$16-17$ & NPL22$_{A}$,22$_{B}$, 29, 30   & 4   \\
$17-18$ & NPL34, 35$_{A}$, 32$_{A}$      & 3   \\
$18-19$ & NPL32$_{B}$, 36                & 2   \\
$19-20$ & NPL37, 38, 39                  & 3   \\
$20-21$ & NPL40                          & 1   \\
$21-22$ & NPL35$_{B}$                    & 1   \\
\hline
\end{tabular}
\end{flushleft}
\end{table}

Figure~\ref{Figimf} suggests a rising MF towards lower masses with a power-law index between 0 and 1, but there are several uncertainties one should be aware of. 

The observed number of objects is small, but it remains clear that the LF drops from $I=15$ to the survey limit. 

We cannot say how many Pleiades BDs that are lost in unresolved binaries. However, we estimate a likely upper limit by assuming that the 30\,\% multiple fraction for M dwarfs \cite{reid97} applies to all our single Pleiades candidates which are brighter than the BD limit at $I\sim17$. If those companions all are below the BD limit, 8 BDs would have been lost. Of course, we do not know their possible distribution, but adding 8 BDs to the LF still leaves it close to the $n=1$ curve. From this crude discussion it follows that lost companions are unlikely to raise the MF index significantly above 1.

There is also a normalization error, which does not affect the falling trend, however. 

The model in Baraffe et al. \cite*{baraffe98} is state-of-the-art and based on the latest generation of non-grey atmospheres and should provide the best possible mass-luminosity relations for the Pleiades BD sequence. Since it has not yet been possible to derive an empirical mass-luminosity relation for substellar objects, we do not try to estimate the uncertainties in mass that may arise from the model.

\section{Summary}
We have performed a deep 850 arcmin$^{2}$ $RIJK$ survey in the central area of the Pleiades cluster.
Photometry in $I$, $J$ or $K$ is presented for 32 previously known and 7 new likely members. Four of those are below the BD limit, and the faintest one would be the lowest mass BD found in the Pleiades so far, if its membership could be confirmed.

The overall agreement in the photometry with other surveys is satisfactory. Teide1 is an exception. We find $I=19.26$, $\sim$\,0.5 mag fainter than ZRM. Based on RGO archive data and a roughly constant $J$ magnitude we conclude that the reason for this discrepancy most likely sits in the photometry of ZRM and not in intrinsic variability.  

A number of very red faint objects were found below the Pleiades sequence. Two of those were measured in $I$, $J$ and $K$ and fit in as GD165B-type objects, possible field BDs.

After splitting PPL15 and two other probable binaries into components, the Pleiades LF was compared to model LFs derived from the most recent theoretical mass-luminosity relations. The observed LF supports an MF-index between 0 and 1. Thus, even if the MF seems to rise for low-mass BDs in the Pleiades, it is not steep enough to leave more than a few percent of the cluster's mass in BDs, which is consistent with dynamical findings \cite{pinfield97b}. 

\begin{acknowledgements}
      I would like to thank the NOT OPC for allocating time for this project 
      which is an essential part in my coming thesis. Thanks also to N.C. Hambly who provided proper motions for some of my candidates and to I. Baraffe who provided the model on ascii format.
      This work was supported by the Nordic Optical Telescpe Scientific 	
      Association and the Swedish Natural Science Research Council (NFR). This research has made use of the RGO La Palma archive.
\end{acknowledgements}


\begin{thebibliography}{}

\bibitem[\protect\astroncite{{Allard} \& {Hauschildt}}{1997}]{allard97}
{Allard} F., {Hauschildt} P., 1997, Model atmospheres: Brown dwarfs from the stellar perspective.
\newblock In: R. {Rebolo}, E. {Martin}, M.~R. {Zapatero Osorio} (eds.), {
  Brown dwarfs and extrasolar planets: ASP Conf. Ser.},
\newblock in press

\bibitem[\protect\astroncite{{Bahcall}}{1986}]{bahcall86}
{Bahcall} J.~N., 1986,
\newblock { ARA\&A} { 24}, 577

\bibitem[\protect\astroncite{{Bahcall} \& {Soneira}}{1980}]{bahcall80}
{Bahcall} J.~N., {Soneira} R.~M., 1980,
\newblock { ApJS} { 44}, 73

\bibitem[\protect\astroncite{{Baraffe} et~al.}{1998}]{baraffe98}
{Baraffe} I., {Chabrier} G., {Allard} F, {Hauschildt} P., 1998,
\newblock { A\&A} { submitted}

\bibitem[\protect\astroncite{{Basri} et~al.}{1996}]{basri96}
{Basri} G., {Marcy} G.~W., {Graham} J.~R., 1996,
\newblock { ApJ} { 458}, 600

\bibitem[\protect\astroncite{{Basri} \& {Martin}}{1997}]{basri97}
{Basri} G., {Martin} E., 1997, Is PPL15 a binary brown dwarf system?
\newblock In: R. {Rebolo}, E. {Martin}, M.~R. {Zapatero Osorio} (eds.), {
  Brown dwarfs and extrasolar planets: ASP Conf. Ser.},
\newblock in press

\bibitem[\protect\astroncite{{Burstein} \& {Heiles}}{1982}]{burstein82}
{Burstein} D., {Heiles} C., 1982,
\newblock { AJ} { 87}, 1165

\bibitem[\protect\astroncite{{Casali} \& {Hawarden}}{1992}]{casali92}
{Casali} M., {Hawarden} M.~T., 1992,
\newblock { JCMT-UKIRT Newsletter} { 4}, 33

\bibitem[\protect\astroncite{{Cossburn} et~al.}{1997}]{cossburn97}
{Cossburn} M.~R., {Hodgkin} S.~T., {Jameson} R.~F., {Pinfield} D.~J.,
  1997,
\newblock { MNRAS} { 288}, L23

\bibitem[\protect\astroncite{{Festin}}{1997}]{festin97a}
{Festin} L., 1997,
\newblock { A\&A} { 322}, 455

\bibitem[\protect\astroncite{{Gould} et~al.}{1997}]{gould97}
{Gould} A., {Bahcall} J.~N., {Flynn} C., 1997,
\newblock { ApJ} { 482}, 913

\bibitem[\protect\astroncite{{Hambly} et~al.}{1991}]{hambly91}
{Hambly} N.~C., {Jameson} R.~F., {Hawkins} M.~R.~S., 1991,
\newblock { MNRAS} { 253}, 1

\bibitem[\protect\astroncite{{Hambly} et~al.}{1993}]{hambly93}
{Hambly} N.~C., {Hawkins} M.~R.~S., {Jameson} R.~F., 1993,
\newblock { A\&AS} { 100}, 607 (HHJ)

\bibitem[\protect\astroncite{{Hunt} et~al.}{1995}]{hunt95}
{Hunt} L.~K., {Migliorini} S., {Testi} L., et al., 1995,
\newblock { AJ},
\newblock submitted

\bibitem[\protect\astroncite{{Jameson} \& {Skillen}}{1989}]{jameson89}
{Jameson} R.~F., {Skillen} I., 1989,
\newblock { MNRAS} { 239}, 247 (JS)

\bibitem[\protect\astroncite{{Jones} et~al.}{1994}]{jones94}
{Jones} H.~R.~A., {Longmore} A.~J., {Jameson} R.~F., {Mountain} C.~M.,
  1994,
\newblock { MNRAS} { 267}, 413

\bibitem[\protect\astroncite{{Landolt}}{1992}]{landolt92}
{Landolt} A.~U., 1992,
\newblock { AJ} { 104}, 340

\bibitem[\protect\astroncite{{Leggett}}{1992}]{leggett92}
{Leggett} S.~K., 1992,
\newblock { ApJS} { 82}, 351

\bibitem[\protect\astroncite{{Mermilliod} et~al.}{1997}]{mermilliod97}
{Mermilliod} J.~C., {Turon} C., {Robichon} N., {Arenou} F., 1997, The distance of the Pleiades and nearby clusters
\newblock In: B. {Battrick} (ed.), { HIPPARCOS VENICE '97}, p. 643, ESA SP-402

\bibitem[\protect\astroncite{{Pinfield} et~al.}{1997}]{pinfield97b}
{Pinfield} D.~J., {Jameson} R.~F., {Hodgkin} S.~T., 1997, The mass of the Pleiades
\newblock In: R. {Rebolo}, E. {Martin}, M.~R. {Zapatero Osorio} (eds.), {
  Brown dwarfs and extrasolar planets: ASP Conf. Ser.},
\newblock in press

\bibitem[\protect\astroncite{{Rebolo} et~al.}{1995}]{rebolo95}
{Rebolo} R., {Zapatero Osorio} M.~R., {Martin} E.~L., 1995,
\newblock { Nat} { 377}, 129

\bibitem[\protect\astroncite{{Rebolo} et~al.}{1996}]{rebolo96}
{Rebolo} R., {Martin} E.~L., {Basri} G., {Marcy} G.~W., {Zapatero
  Osorio} M.~R., 1996,
\newblock { ApJ} { 469}, L53

\bibitem[\protect\astroncite{{Reid} \& {Gizis}}{1997}]{reid97}
{Reid} I.~N., {Gizis} J.~E., 1997,
\newblock { AJ} { 113}, 2246

\bibitem[\protect\astroncite{{Santiago} et~al.}{1996}]{santiago96}
{Santiago} B.~X., {Gilmore} G., {Elson} R.~A.~W., 1996,
\newblock { MNRAS} { 281}, 871

\bibitem[\protect\astroncite{{Scalo}}{1986}]{scalo86}
{Scalo} J.~M., 1986,
\newblock { Fund. Cosm. Phys.} { 11}, 1

\bibitem[\protect\astroncite{{Schultz}}{1997}]{schultz97}
{Schultz} G., 1997, Searching for new brown dwarf candidates in a Pleiades IJK imaging survey
\newblock In: R. {Rebolo}, E. {Martin}, M.~R. {Zapatero Osorio} (eds.), {
  Brown dwarfs and extrasolar planets: ASP Conf. Ser.},
\newblock in press

\bibitem[\protect\astroncite{{Stauffer} et~al.}{1989}]{stauffer89}
{Stauffer} J., {Hamilton} D., {Probst} R., {Rieke} G., {Mateo} M.,
  1989,
\newblock { ApJ} { 344}, L21

\bibitem[\protect\astroncite{{Stauffer} et~al.}{1994a}]{stauffer94a}
{Stauffer} J.~R., {Hamilton} D., {Probst} R.~G., 1994a,
\newblock { AJ} { 108}, 155

\bibitem[\protect\astroncite{{Stauffer} et~al.}{1994b}]{stauffer94b}
{Stauffer} J.~R., {Liebert} J., {Giampapa} M., et al., 1994b,
\newblock { AJ} { 108}, 160

\bibitem[\protect\astroncite{{Stauffer} et~al.}{1995}]{stauffer95}
{Stauffer} J.~R., {Liebert} J., {Giampapa} M., 1995,
\newblock { AJ} { 109}, 298

\bibitem[\protect\astroncite{{Steele} \& {Jameson}}{1995}]{steele95}
{Steele} I.~A., {Jameson} R.~F., 1995,
\newblock { MNRAS} { 272}, 630

\bibitem[\protect\astroncite{{Steele} et~al.}{1993}]{steele93}
{Steele} I.~A., {Jameson} R.~F., {Hambly} N.~C., 1993,
\newblock { MNRAS} { 263}, 647

\bibitem[\protect\astroncite{{Willliams} et~al.}{1996}]{williams96}
{Willliams} D.~M., {Boyle} R.~P., {Morgan} W.~T., et al., 1996,
\newblock { ApJ} { 464}, 238

\bibitem[\protect\astroncite{{Winkler}}{1997}]{winkler97}
{Winkler} H., 1997,
\newblock { MNRAS} { 287}, 481
 
\bibitem[\protect\astroncite{{Zapatero Osorio}}{1997}]{zapatero97c}
{Zapatero Osorio} M.~R., 1997, Revealing the brown dwarf population in the Pleiades cluster
\newblock In: R. {Rebolo}, E. {Martin}, M.~R. {Zapatero Osorio} (eds.), {
  Brown dwarfs and extrasolar planets:ASP Conf. Ser.},
\newblock in press

\bibitem[\protect\astroncite{{Zapatero Osorio} et~al.}{1997a}]{zapatero97a}
{Zapatero Osorio} M.~R., {Rebolo} R., {Martin} E.~L., 1997a,
\newblock { A\&A} { 317}, 164 (ZRM)

\bibitem[\protect\astroncite{{Zapatero Osorio} et~al.}{1997b}]{zapatero97b}
{Zapatero Osorio} M.~R., {Martin} E.~L., {Rebolo} R., 1997b,
\newblock { A\&A} { 323}, 105 (ZMR)

\end{thebibliography}
\end{document}